\documentclass[preprint,aps]{revtex4}

\usepackage{graphicx}
\usepackage{dcolumn}
\usepackage{bm}

\begin{document}
\newcommand{\be}{\begin{equation}}   \newcommand{\ee}{\end{equation}}
\newcommand{\mean}[1]{\left\langle #1 \right\rangle}
\newcommand{\abs}[1]{\left| #1 \right|}
\newcommand{\la}{\langle}
\newcommand{\ra}{\rangle}
\newcommand{\lb}{\left(}
\newcommand{\rb}{\right)}
\newcommand{\norm}[1]{\left\|#1\right\|}
\newcommand{\RA}{\rightarrow}
\newcommand{\tet}{\vartheta}
\newcommand{\eps}{\varepsilon}

\preprint{APS/123-QED}

\title{Noise level estimation of time series using coarse grained entropy}

\author{Krzysztof Urbanowicz}
 \email{urbanow@if.pw.edu.pl}
\affiliation{Faculty of Physics and Center of Excellence for Complex Systems
Research \\Warsaw University of Technology\\ Koszykowa 75, PL--00-662 Warsaw,
Poland}
 \affiliation{Institute for Economics and Traffic\\ Dresden University of
  Technology \\Andreas-Schubert-Strasse 23, D-01062 Dresden, Germany}.
\author{Janusz A. Ho{\l}yst}%
 \email{jholyst@if.pw.edu.pl}
\affiliation{Faculty of Physics and Center of Excellence for
Complex Systems Research \\Warsaw University of Technology\\
Koszykowa 75, PL--00-662 Warsaw, Poland}
 \affiliation{Institute for Economics and Traffic\\ Dresden University of
  Technology \\Andreas-Schubert-Strasse 23, D-01062 Dresden, Germany}.

\date{\today}

 \begin{abstract}
We present a method of noise level estimation that is valid even
for high noise levels. The method makes use of the functional
dependence of coarse grained correlation entropy $K_2(\eps)$ on
the threshold parameter $\eps$. We show that the function
$K_2(\eps)$  depends in a characteristic way on the noise standard
deviation $\sigma$. It follows that observing  $K_2(\eps)$ one can
estimate the noise level $\sigma$. Although the theory has been
developed for the gaussian noise added to the observed variable we
have checked numerically that the method is also valid  for the
uniform noise distribution and for the case of Langevine equation
corresponding to the dynamical noise.  We have verified the
validity of our method by applying it to estimate the noise level
in several  chaotic systems and in the Chua electronic circuit
contaminated by noise.
\end{abstract}
\maketitle

\section{Introduction}
 \par  It is a common case that observed data  are contaminated
 by a
 noise (for a review of methods of
 nonlinear time series analysis see
 \cite{kantzschreiber,abarbanel}). The presence of noise can substantially affect invariant system parameters as a dimension,
     entropy or Lyapunov exponents. In fact Schreiber
     \cite{Schreiber1} has shown that even $2\%$ of noise can make a dimension
     calculation misleading. It follows that the assessment of the noise level can
     be crucial    for
     estimation of system invariant parameters.
        Even
        after performing a noise reduction one is interested to evaluate the noise
        level in the cleaned data.
In the experiment the noise is often regarded
 as a measurement uncertainty which corresponds to a random variable added to
 the system temporary state or to the experiment outcome. This kind of noise is
 usually called  the {\it measurement} or the {\it additive} noise.
 Another case is the noise influencing the system dynamics, what corresponds to the Langevine
 equation and can be called the {\it dynamical} noise. The second case is more difficult to
 analyze because the
 noise acting at moment
 $t_0$
 usually changes the trajectory for $t>t_0$.
 It follows that there is no clean trajectory and instead of it an $\epsilon$-shadowed
 trajectory occurs \cite{Farmer}. For real data a signal (e.g. physical experiment data
    or economic data) is subjected to the mixture of both
    kinds of noise (measurement and dynamical).

    \par

   \par
 Schreiber has developed a method of noise level estimation \cite{Schreiber1}
  by evaluating the influence of noise on the correlation
 dimension
 of  investigated system.
The Schreiber method is valid for
 rather small
 gaussian measurement
 noise and needs values of the embedding dimension $d$, the embedding delay $\tau$ and
  the characteristic dimension $r$ spanned by the system dynamics.
 \par Diks \cite{Diks} investigated properties of correlation integral with the gaussian kernel in the
  presence of noise.  The Diks method makes use of a fitting function for correlation integrals
   calculated from
 time series for different thresholds $\eps$. The
 function depends on system variables $K_2$ (correlation entropy), $D_2$ (correlation dimension), $\sigma$ (standard noise deviation)
  and a normalizing constant $\Phi$. These four variable are estimated
  using the least squares fitting. The Diks method \cite{Yu} is valid for a noise level up
 to $25\%$ of signal variance and
  for various measurement noise distributions.
  The Diks's method needs optimal values of the embedding dimension $d$,
   the embedding delay $\tau$ and the maximal threshold $\eps_c$.
 \par
 Hsu et al. \cite{Hsu} developed a method of noise \textit{reduction} and they used this method for noise level estimation. The method explored the
 local-geometric-projection principle  and is useful for various noise distributions but rather small noise
 levels. To use the method one needs to choose a number of
 neighboring points to be regarded, an appropriate number of iterations as well as  optimal parameters
 values $d$ and $\tau$.
\par
 Oltmans et al. \cite{Oltmans} considered influence of noise on
 the probability density function $f_n(\eps)$ but they could take into account only a small measurement
 noise. They used a fit of $f_n(\eps)$ to
the corresponding function which was found for small $\eps$. Their fitting
function is similar to the
 probability density distribution that we receive from correlation
 integrals
 $\frac{1}{N^2} DET_n(\eps)$. The method needs as input parameters values of $d$,
 $\tau$ and
 ${\eps}_c$.

\par Our method has its origin in recurrence plots (RPs)
\cite{Eckmann} and it uses RPs quantities to characterize the
data. Recurrence plots were originally introduced by Eckmann
\cite{Eckmann} as a useful graphic way for data analysis. The plot
is defined as a matrix $N$ x $N$ where a dot (i, j) is drawn when
   $\|\vec{y}_i-\vec{y}_j\|<\varepsilon$
    ($\varepsilon$ is a given threshold). By recurrence plots one can study data
    stationarity \cite{Zbilut,manetti,giuliani},
    as well as their recurrence and deterministic properties \cite{Holyst,Atay,Chai}.
    The approach was also applied for parameter
    optimizing \cite{Matassini} in the local projection method of noise reduction \cite{Hegger}.
     RPs can be easy used  to calculate
    characteristic system parameters like the correlation entropy \cite{Faure}, what will be
    performed
    in our case. Lines of black dots parallel to the main diagonal can appear in recurrence
    plots
    and their number can serve as a measure of determinism \cite{Zbilut}.
    In our method we take into account a number of lines
     $DET_n$ of the length $n$ or longer by the embedding dimension  $d=1$.
     We use the fact that there is a straightforward
     relation between $DET_n$ and the correlation integral \cite{Faure}.

\par The crucial point of our method is fitting of a proper function to the estimated correlation
entropy $K_2$. In fact  similar considerations can be performed
for Kolmogorov-Sinai entropy \cite{Shan,wang,zyczkowski} $K_1$
using for example the approach given in  \cite{cohen} but in such
a case a much larger number of data is needed since the $K_1$
entropy is more sensitive to regions of the phase space with small
values of invariant measure.  The method is not too time
consuming, e.g. a calculation of entropy for 100 various
thresholds and $N=3000$ data points needed a few minutes
\cite{computer}. Our method does not demand any input parameters
like the embedding dimension d or the embedding delay $\tau$. The
minimal and maximal values of the threshold parameter $\epsilon$
can be automatically estimated. In all considerations we use the
maximum norm to save the computation time and to perform analytic
expansions. It is known that  in the limit $\epsilon \RA 0$ the
behavior of invariant system parameters does not depend on the
type of used norm. In our case features of coarse grained entropy
are considered and the value of the threshold parameter $\eps$
should be comparable to the noise level. It follows that one can
not exclude that the type of applied norm affects the functional
dependence of the coarse grained entropy $K_2(\eps)$ in the
presence of noise of a large or medium value.
\par We stress here that our method is provided for a noise level
estimation.
 The method is not equivalent to noise filters
  that allow to extract an original non disturbed signal from noisy time series
    \cite{Schreiber,Farmer,Saure}.

\section{Entropy estimation for a time series in the noise absence}
\par
Let $\{x_i\}$ where $i=1,2,...,N$ be a time series and
$\vec{y}_i=\{x_i,x_{i+\tau},...,x_{i+(n-1)\tau}\}$ a corresponding
$n$-dimensional vector constructed in the embedded space  where
$n$ is an embedding dimension and $\tau$ is an embedding delay.
The correlation integral calculated in the embedded space
$\vec{y}_i$ is \be \label{eq.1a} C^n(\eps)=\frac{1}{N^2} \sum_i^N
\sum_{j\neq i}^N\theta(\eps-\norm{\vec{y}_i-\vec{y}_j})\ee where
$\theta$ is the Heavyside step function. If $\norm{...}$ is the
maximum norm the correlation integral $C^n(\eps)$ is proportional
to the number $DET_n(\eps)$ of lines of the length $n$ in the RP
constructed from the data set $\{x_i\}$  \cite{Faure}
\begin{eqnarray}  C^n(\eps)=\frac{1}{N^2}\sum_i \sum_{j\neq
i}\theta(\eps-\abs{x_{i}-x_j})\theta(\eps-\abs{x_{i+\tau}-x_{j+\tau}})\nonumber\\
 ...\theta(\eps-\abs{x_{i+(n-1)\tau}-x_{j+(n-1)\tau}})=\frac{DET_n(\eps)}{N^2}
 \label{eq.1}.\end{eqnarray} The correlation entropy  \cite{Proccacia,Benettin} can now be
calculated as\be K_2=\lim_{\eps\rightarrow
0}\lim_{n\rightarrow\infty}\ln\frac{DET_n(\eps)}{DET_{n+1}(\eps)}\approx-\frac{d\ln(DET_n(\eps))}{dn}\label{eq.2}.\ee
We assume that Eq.~(\ref{eq.2}) is approximately valid for $n\geq
2$ thus \be DET_n=DET_2 e^{-(n-2)K_2}\label{eq.DET2}.\ee  Let us
introduce the following convention for lines counting: if there is
a line of the length $n$ then it includes one line of the length
$n-1$, one line of the length $n-2$ etc. Using Eq.~(\ref{eq.DET2})
one can easy find the average line length $\mean{n}$
\begin{eqnarray}  \mean{n}=\frac{\sum_{n=2}^{\infty}\lb
DET_n+DET_{n+2}-2DET_{n+1}\rb \cdot n}{\sum_{n=2}^{\infty}\lb
DET_n+DET_{n+2}-2DET_{n+1}\rb} \nonumber\\ \cong
\frac{\sum_{n=2}^{\infty}n \cdot e^{-(n-2)
K_2}}{\sum_{n=2}^{\infty}e^{-(n-2) K_2}
}=\frac{2-e^{-K_2}}{1-e^{-K_2}}\;. \end{eqnarray} The above
formula neglects all lines of the length $n=1$. Now the entropy
can be approximated as \be
K_2\approx\ln\frac{\mean{n}-1}{\mean{n}-2}\label{eq.3}.\ee \par
The relation between the entropy, dimension and correlation
integral is given by the well known formula
\cite{Schuster,Grassberger} \be\lim_{n\RA\infty}\lim_{\eps\RA
0}\ln\frac{1}{N^2} DET_n\lb\eps\rb=D_2\ln\eps-n\tau
K_2\label{eq.ci}\ee
 thus the logarithm of the correlation integral is a linear function
 of entropy $K_2$ and system dimension $D_2$. On the other hand the correlation dimension $D_2$
 is independent of the embedding dimension $d$ if
 the latter is large enough. We use this fact and in the next section we will estimate the noise effect on the
 dimension $D_2$
 as well as
 on the length $n$ of the line in RP where the line length  corresponds to the embedding dimension.
 At the end we will incorporate  both effects into Eq.~(\ref{eq.1}) to
 reproduce the
 complete influence of noise on the correlation integral.

\section{Influence of noise on
correlation integral}
\par Let us modify the definition of $DET_n$ in such a way that the influence of noise on entropy can
 be analytically estimated. First we change Eq.~(\ref{eq.1a}) to
the equivalent form \be  DET_n (\eps)=\sum_i^N\sum_{i \neq j}^N
\theta\lb\sum_{k=0}^l
\theta\lb\eps-\abs{x_{i+k}-x_{j+k}}\rb-n\rb\label{eq.dat}\ee where
$l$ is the length of the recurrence line beginning at the point
$(i,j)$. Eq.~(\ref{eq.dat}) is valid provided that one assumes
$\theta(0)=1$ for the Heavyside function. The function $\theta$ in
Eq.~(\ref{eq.1a}) is called a kernel function \cite{Ghez} and it
can be written in a general way  as $\rho_{\eps}\lb r\rb$. Now let
us use the fact \cite{Ghez} that the kernel function can be
replaced by any monotonically decreasing function $\rho_{\eps}(r)$
with a bandwidth $\eps$ such that $\lim_{r\rightarrow
0}r^{-p}\rho_{\eps} (r)=0$ for $\eps>0$ and any $p\geq 0$. The
bandwidth $\eps$ of the kernel function corresponds to the
threshold $\eps$. It follows that we can replace the inner
$\theta\lb\eps-r\rb$ function in Eq.~(\ref{eq.dat}) by a new
linear continuous function
\begin{equation} \theta(\eps-r)\Rightarrow \rho_{\eps}(r)=\left\{
\begin{array}{ccc}
\frac{\eps-r}{\eps}& \mbox{for} &0\leq r\leq \eps \\
 0 & \mbox{for} &r>\eps \end{array}\right\}\end{equation} and simultaneously we lower the threshold
in outer $\theta$ function
  by the constant $\beta=\frac{1}{\sqrt{\pi}}$. We have checked that other choices of $\beta$ bring similar
  results. Now instead of Eq.~(\ref{eq.dat}) we have \be DET'_n (\eps)=\sum_i^N\sum_{i
\neq j}^N \theta\lb\sum_{k=0}^n
\frac{\eps-\abs{x_{i+k}-x_{j+k}}}{\eps}-\beta n\rb\label{eq.4}.\ee
We use the above expression to calculate the mean line length
$\mean{n}$. Practically the length of each line is calculated as
the maximal value of the parameter $n$ in Eq.~(\ref{eq.4})
provided that the $\theta$ function equals to $1$. Having
$\mean{n}$ we calculate the system entropy $K_2$ using the
Eq.~(\ref{eq.3}). \par Now let us consider the influence of
uncorrelated gaussian noise $\eta_{i}$ added
 to the observed system variable $x_{i}$. The equation
 (\ref{eq.4}) is replaced by the following approximation
 \begin{widetext}
\begin{eqnarray} DET'_n (\eps)=\sum_i^N\sum_{i \neq j}^N
\theta\lb\sum_{k=0}^n
\frac{\eps-\abs{x_{i+k}+\eta_{i+k}-x_{j+k}-\eta_{j+k}}}{\eps}-\beta
n\rb\nonumber\\ \cong\sum_i^N\sum_{i \neq j}^N
\theta\lb\sum_{k=0}^n
\frac{\eps-\abs{x_{i+k}-x_{j+k}}}{\eps}-n\frac{\sqrt{\alpha^2\eps^2+2\sigma^2}-\alpha\eps}{\eps}-\beta
n\rb\ \label{wide}\end{eqnarray} \end{widetext} where $\sigma$ is
the standard noise deviation and $\alpha$  is a constant of order
of $1$ that depends on the distribution of  $\abs{x_i-x_j}$. One
can easily derive Eq.~(\ref{wide}) assuming that
$\sigma_{x}\approx \alpha\eps$ where $\sigma_{x}$ stands for a
standard deviation of $\abs{x_i-x_j}\in \lb 0,\eps\rb$. When the
differences  $\abs{x_i-x_j}$ are uniformly distributed in the
region $\lb 0,\eps\rb$ then $\alpha = \frac{1}{\sqrt{3}}$.

Comparing Eq.~(\ref{wide})  to Eq.~(\ref{eq.dat}) and
Eq.~(\ref{eq.4}) we see that the effect of noise corresponds
formally to the change
\begin{equation} n\RA n\lb
1+\sqrt{\pi}\frac{\sqrt{\eps^2/3+2\sigma^2}-\eps/\sqrt{3}}{\eps}\rb\;.
\end{equation} Instead of the second part of lhs Eq.~(\ref{eq.ci}) we
have \be -n\tau K_2\rightarrow-n \tau K_2(\eps)\lb
1+\sqrt{\pi}\frac{\sqrt{\eps^2/3+2\sigma^2}-\eps/\sqrt{3}}{\eps}\rb\label{eq.9}.
\ee For a small noise ($\sigma\ll\epsilon$) the last equation can
be transformed to
\begin{equation} -n\tau K_2\rightarrow-n \tau K_2(\eps)\lb
1+\sqrt{3\pi}\frac{\sigma^2}{\eps^2}\rb \end{equation} what is in
agreement with the well known result \cite{boffetta,cencini} for
the noise entropy in the case of noise spectrum
$S(\omega)\sim\omega^{-2}$ \begin{equation} K_{noisy}\sim
\frac{1}{\eps^2}\;.\end{equation} The Eq.~(\ref{eq.9}) expresses
the influence of noise on the line length $n$.  On the other hand
Schreiber has shown \cite{Schreiber1} that the influence of noise
can be described by the substitution  in the equation
(\ref{eq.ci}) \be D_2\rightarrow\lb D_2+(n-r)g\lb
\frac{\eps}{2\sigma}\rb\rb \label{eq.1w}\ee where\begin{equation}
g(z)=\frac{2}{\sqrt{\pi}}\frac{ze^{-z^2}}{erf(z)}\end{equation}
and the parameter $r$ follows from the method of singular value
decomposition used in \cite{Schreiber1}.
\par Combining
Eq.~(\ref{eq.ci}) with results (\ref{eq.9}) and (\ref{eq.1w}) we
get
\begin{eqnarray} DET_n(\eps)\sim \eps^{\lb D_2+(n-r)g\lb
 \eps/2\sigma \rb\rb}
 \nonumber\\\times
\exp \lb -n \tau K_2(\eps)\lb
1+\sqrt{\pi}\frac{\sqrt{\eps^2/3+2\sigma^2}-\eps/\sqrt{3}}{\eps}\rb\rb
\end{eqnarray} where $K_2(\eps)$ is the coarse grained entropy of
the clean signal. The explicit form of the function $K_2(\eps)$ is
unknown. A good fit that seems to be valid for several systems is
\begin{equation} K_2(\eps)=\kappa+b \ln \lb
1-a\eps\rb\end{equation} where the constant $\kappa$ corresponds
to the correlation entropy while the second term describes the
effect of the coarse graining. We stress here that the precise
value of the latter function is not needed for our approach of
noise level estimation because we are left with some free
parameters. It follows that one can estimate the coarse grained
entropy of the signal with noise as
\begin{eqnarray} K_{noisy}(\eps)=-\frac{d \ln \lb
DET_n(\eps)\rb}{dn} \nonumber\\=-\frac{1}{\tau}g\lb\frac{\eps}{2
\sigma}\rb\ln{\eps}+K_2(\eps)\lb
1+\sqrt{\pi}\frac{\sqrt{\frac{\eps^2}{3}+2\sigma^2}-\frac{\eps}{\sqrt{3}}}{\eps}\rb
\end{eqnarray} where the function $g(.)$ corresponds to the
influence of noise on the  correlation dimension  while the second
term can be split into the coarse grained entropy of the clean
signal $K_2(\eps)$ and the linear increase of this entropy due to
the presence of the external noise
$\sqrt{\pi}\lb\frac{\sqrt{\eps^2/3+2\sigma^2}-\eps/\sqrt{3}}{\eps}\rb
K_2(\eps)$. To estimate the noise level $\sigma$ one can use the
above dependence of the correlation entropy $K_{noisy}(\eps)$ as
the function of the threshold $\eps$. However we have found that
because of a peculiar behavior of $K_{noisy}(\eps)$ it is more
convenient to fit the function $K_{noisy}(\eps)\cdot \eps^p$
instead of $K_{noisy}(\eps)$ to corresponding experimental data
($p$ is a constant of order of $1$, see next section for
discussion). It follows that we need to estimate five free
parameters $\kappa$, $\sigma$, a, b and c for the function
\begin{widetext}
\begin{eqnarray} K_{noisy}(\eps)\cdot\eps^p=-c\eps^p
g\lb \frac{\eps}{2\sigma}\rb\ln\eps+\lb\kappa+b \ln \lb
1-a\eps\rb\rb\eps^p\lb
1+\sqrt{\pi}\frac{\sqrt{\eps^2/3+2\sigma^2}-\eps/\sqrt{3}}{\eps}\rb\label{eq.13}.\end{eqnarray}
\end{widetext}
The parameter $c$ ($c$ ranges typically  from $0.5$ to $0.7$) has
been introduced for a better agreement to numerical data. To fit
the above function we have used Levenberg-Marquardt method
\cite{NR}. We stress here  we do not need to assume any input
value for the above coefficients  but they  appear as a result of
application of our method.

\section{Noise level estimation: examples}
\par
In practice, all input parameters of the method can be default.
The character of the method causes that the evaluation of the
embedding dimension that usually appears in nonlinear time series
analysis is not needed at all. Since in RP we consider lines of
all lengths larger than $2$, the applied here embedding dimension
is practically the highest as possible for given time series.
\par The first point is to calculate the average line length
$\mean{n}$ for a given threshold and then to find the
corresponding entropy $K_2 (\eps)$ using the formula (\ref{eq.3}).
Having values of entropies for about $100$ different thresholds,
one should rescale the $\eps$-axis. In such a way different
systems with different sizes of attractors can be compared.
Practically we do this by multiplying $\eps$ by some constant
$\gamma$, such that
$\eps_{max}\cdot\gamma=\overline{\eps}_{max}=0.7$ ($\eps_{max}$
has been chosen using the condition $K\lb\eps_{max}\rb=0.015$).
After finding the noise level $\overline{\sigma}$ in the rescaled
data, the corresponding noise of the original time series can be
calculated as $\sigma=\overline{\sigma}/\gamma$.
\par
\begin{figure}
\includegraphics[scale=0.35,angle=-90]{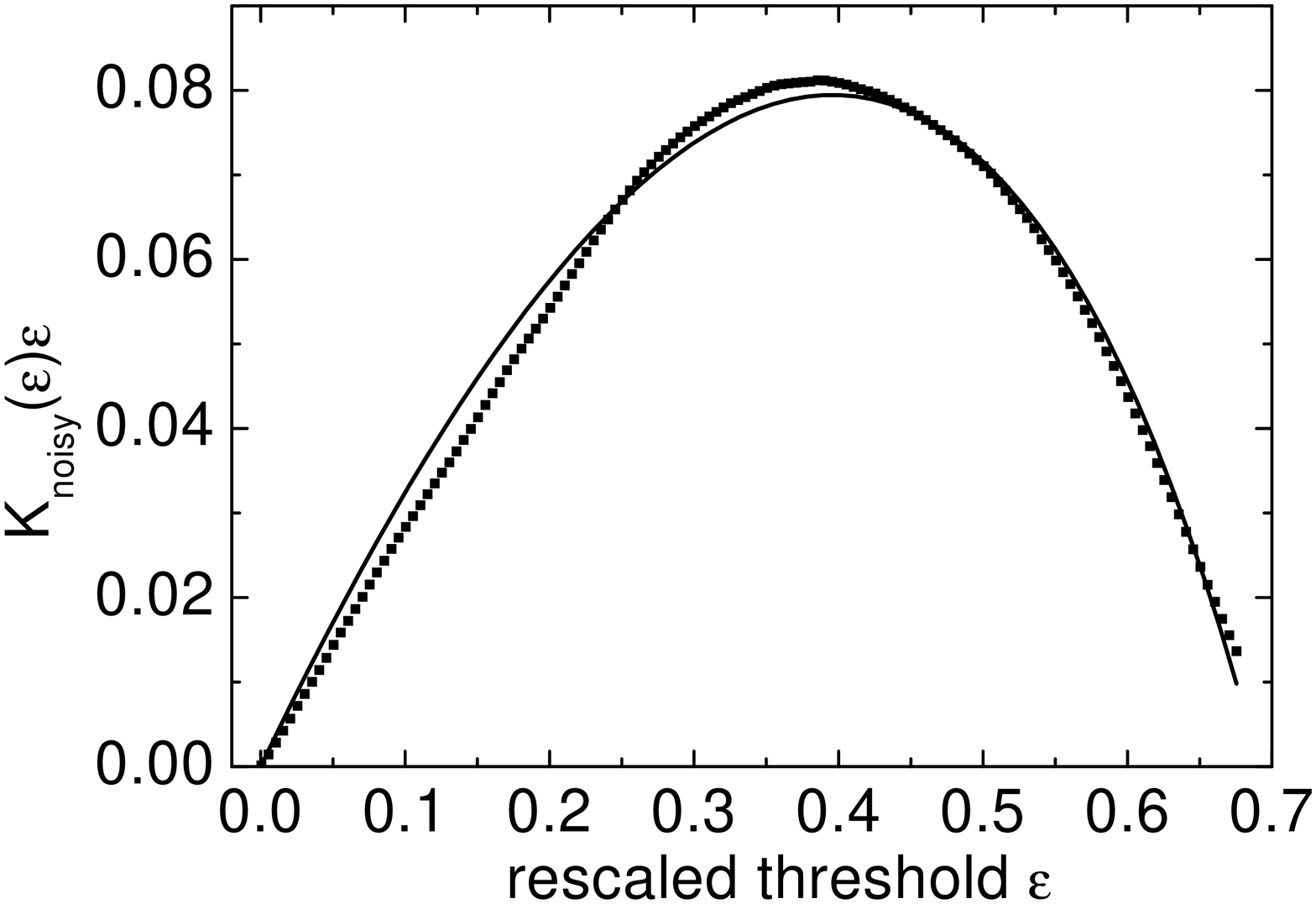}
\caption{\label{fig:fig.1}Chaotic Henon map without a noise. Plot
of coarse grained entropy multiplied by the threshold $\eps$
(squares) calculated from time series  and the fitting function
(\ref{eq.13}) with $p=1$ (line).}
\end{figure}

    One can see the behavior of the fitting
    function (\ref{eq.13}) for the clean signal in Fig.~\ref{fig:fig.1}. For a small threshold
    $\eps << \eps_{max}$ the dependence is linear since for small
    $\eps$ $K_2(\eps)$ is constant.

\begin{figure}
\includegraphics[scale=0.35,angle=-90]{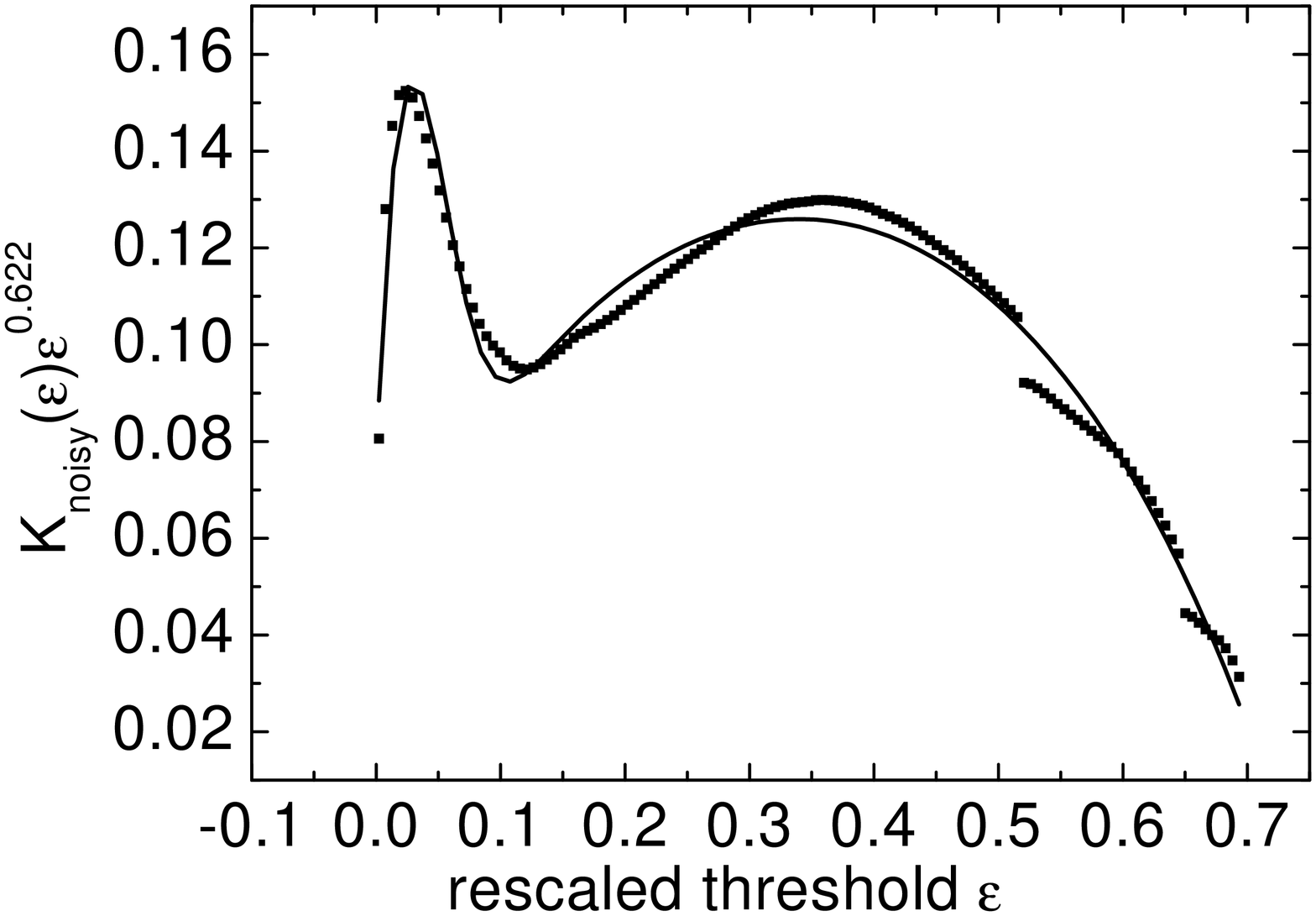}
\caption{\label{fig:fig.2}Chaotic Henon map with measurement noise
$NTS\approx 10\%$. Plot of coarse grained entropy calculated from
the time series multiplied by $\eps^{0.622}$ (squares) and the
fitting function (\ref{eq.13}) with $p=0.622$ (line).}
\end{figure}

The important feature of the plot $K_{noisy}\cdot\lb
\eps\rb\eps^p$ for noisy data is the appearance of two maxima (see
Fig.~\ref{fig:fig.2}). This feature  is helpful for the noise
estimation since
 origins of these maxima are related to the first and second part of rhs of
Eq.~(\ref{eq.13}, i.e. the first maximum is connected to the noise
level, while the second maximum to the finiteness of the
attractor. For a high noise level both maxima merge. The position
of the first peak or the single maximum can be used for additional
noise estimation because one can find that for \be p\simeq
3.441717-\frac{1}{\ln(\sigma)}\label{eq.p}\ee the maximum of
$K_{noisy}\cdot\lb \eps\rb \eps^{p}$ appears at $\eps=\sigma$. The
relation (\ref{eq.p}) gives us the second way, beside
Eq.~(\ref{eq.13}), for estimation of  noise level and for the
control of  results received due to the fitting (\ref{eq.13}).
\par Let us define the percent of noise as the ratio of $\sigma$ to the
standard deviation of data \begin{equation}
\%NTS=\frac{\sigma}{\sigma_{DATA}}\cdot 100\%\end{equation}
\par The estimated values of the standard deviation $\sigma$ received by an
appropriate fit to Eq.~(\ref{eq.13}) for several systems and noise
levels are presented in the table \ref{tab:tab.1}. One can see a
fairly good agreement between the estimated and known level of
noise.
\begin{table}
\caption{\label{tab:tab.1} Results of noise level estimation for
systems with the measurement noise.}
\begin{ruledtabular}
\begin{tabular}{cccc}
System&$\%NTS$&$\sigma$&estimated $\sigma$\\
\hline

  Henon & $0\%$& 0  & $-0.0023\pm 0.0001$ \\
  Henon &$9\%$& 0.1  & $0.1\pm 0.0007$ \\
  Duffing oscylator &$20\%$& 0.4 & $0.46\pm 0.005$\\
  Duffing oscylator &$55\%$& 2 & $1.9\pm 0.02$\\
  Ikeda &$10\%$& 0.07 & $0.07\pm 0.0005$ \\
  Lorenz &$22\%$& 2.2 & $2.2\pm 0.01$ \\
  Roessler &$4\%$& 0.58 & $0.58\pm 0.012$\\
  Roessler &$14\%$& 2 & $1.75\pm 0.01$\\
  Roessler &$35\%$& 6 & $6.16\pm 0.2$\\
  Roessler &$48\%$& 10 & $8.94\pm 0.1$\\
\end{tabular}
\end{ruledtabular}
\end{table}
\par

\par
\begin{figure}
\includegraphics[scale=0.35,angle=-90]{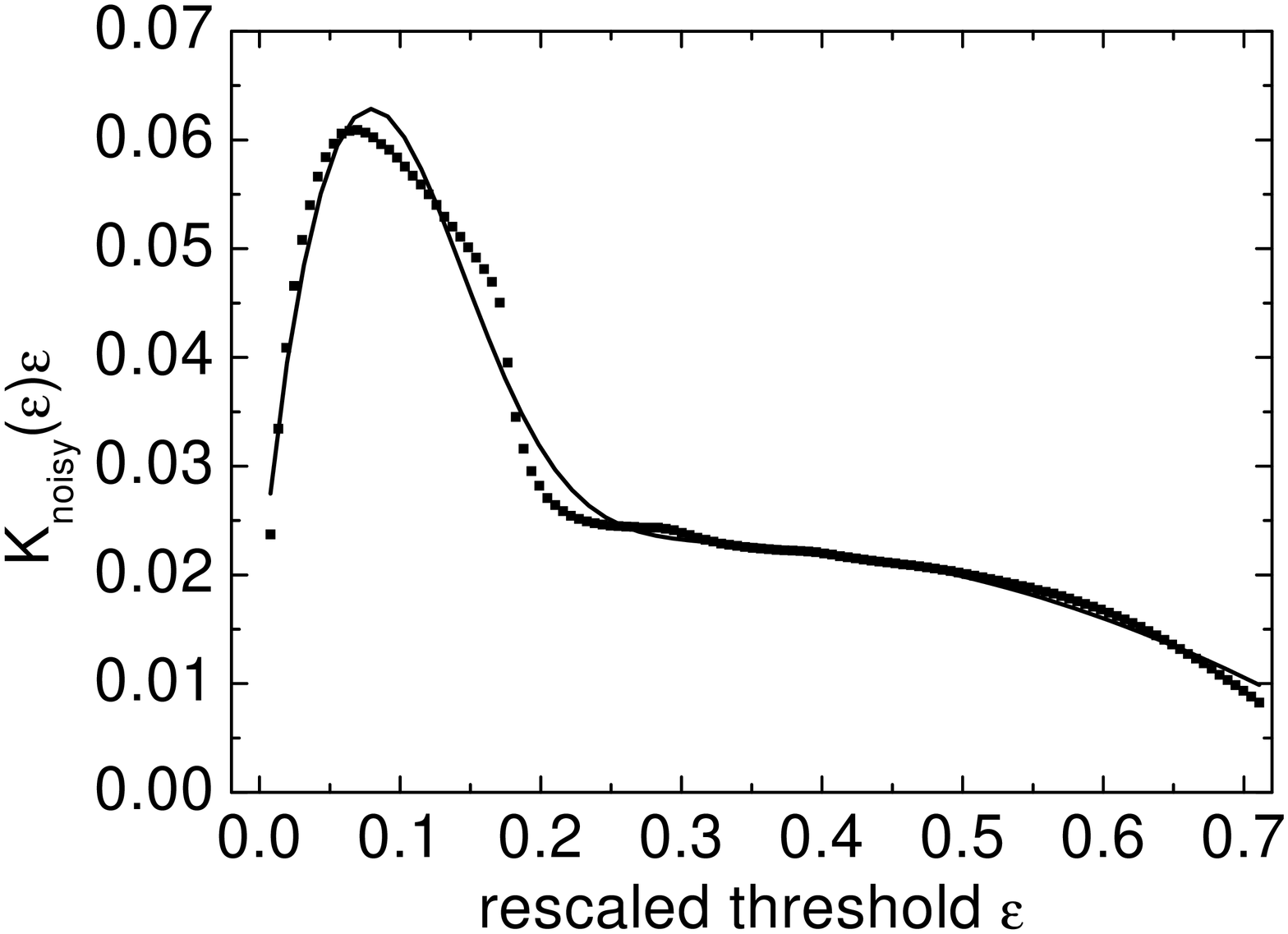}
\caption{\label{fig:fig.3}Chaotic Lorenz model with the dynamical
noise. Plot of coarse grained entropy calculated from the time
series multiplied by the threshold $\eps$ (squares) and the
fitting function (\ref{eq.13}) with $p=1$ (line).}
\end{figure}

 We apply this method for chaotic differential equations
 where the noise $\vec{\eta}_{n}$ is added to system states $\vec{y}_n$,
 calculated by the fourth order Runge-Kutta algorithm.
 It follows that next points of the trajectory are depended in a
nonlinear way on previous noisy contributions \cite{Jaeger} (we
call this kind of noise \textit{a dynamical noise}). In fact we
consider a noise added to the nonlinear map resulting from the
original differential equations and the Runge-Kutta procedure
$\vec{y}_{n+1}=F\lb \vec{y}_{n}+\vec{\eta}_n\rb$. We have found
that the noise level estimated by our method corresponds to the
standard deviation of the noise existing in the system
$\sigma=\sqrt{\mean{\eta^2_n}}$. Fig.~\ref{fig:fig.3} shows that
the behavior of the coarse grained entropy  is similar in the
presence of dynamical and additive noise.

Results for the dynamical noise and a mixture of two kinds of
noise are presented in tables \ref{tab:tab.2} and \ref{tab:tab.3}.
In the table \ref{tab:tab.2} the first three examples correspond
the noise added after writing the value of a variable into a file
and the next examples correspond to the noise added just before
writing a variable to a file.
\par
\begin{figure}
\includegraphics[scale=0.35,angle=-90]{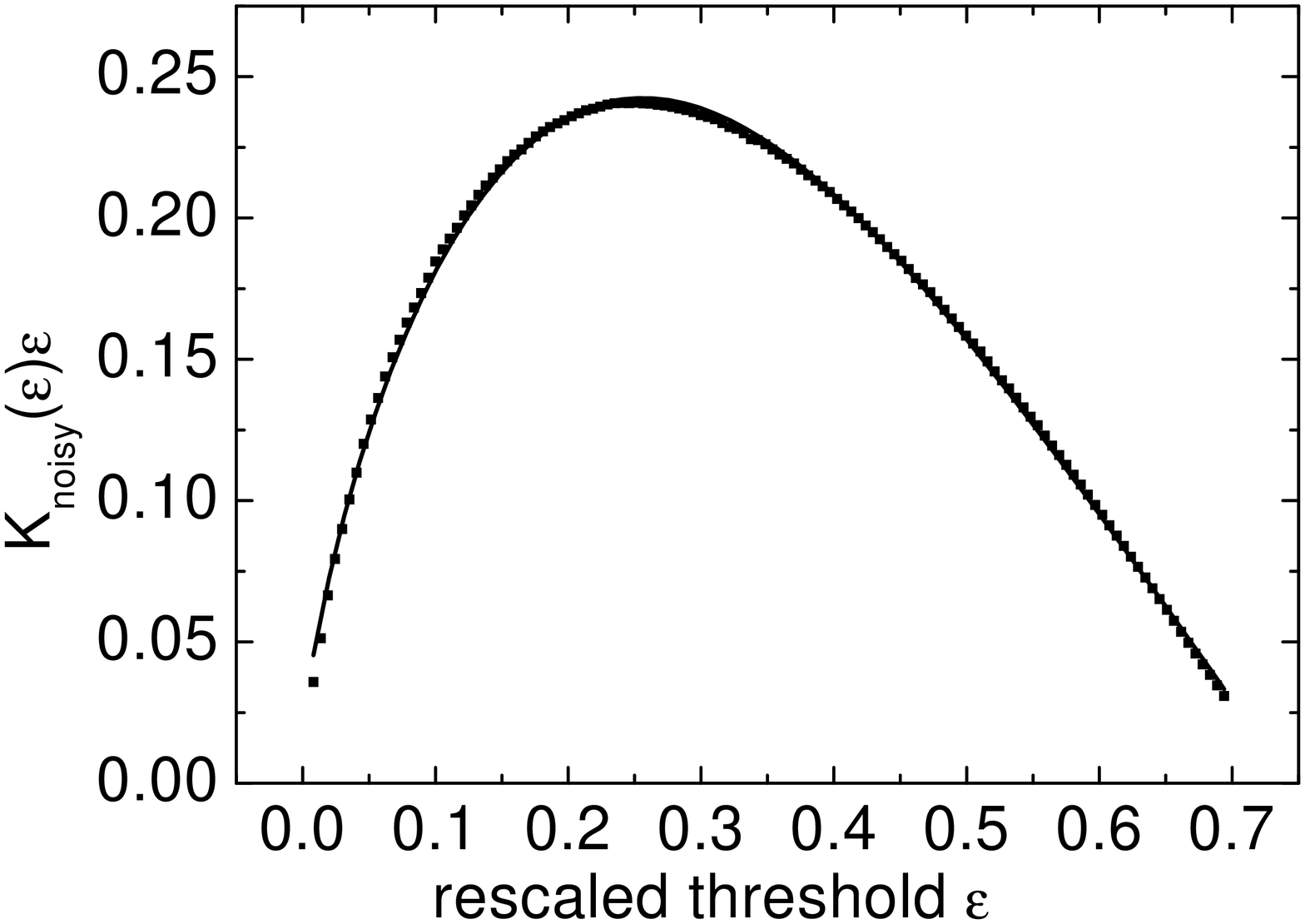}
\caption{\label{fig:fig.4}Noise $\% NTS\approx 100\%$. Plot of
coarse grained entropy calculated from the time series multiplied
by the threshold $\eps$ (squares) and the fitting function
(\ref{eq.13}) with $p=1$ (line).}
\end{figure}

Our method can be useful for evaluation of very high noise levels.
Fig.~\ref{fig:fig.4} shows the plot of the function (\ref{eq.13})
for the noise ($\%NTS\approx 100\%$, $p=1$). In such a case the
error of the estimation is large because we are free to use five
parameters to fit a simple curve.
We have found that for high noise levels it is better to use as
the fitting function a sum of the equations (\ref{eq.13}) with
different exponents $p$ (we have used $p_1=0.5$ and $p_2=7$). It
follows we fit the function
$K_{noisy}\lb\eps\rb\lb\eps^{p_1}+\eps^{p_2}\rb$. The estimation
works better because for different values of $p$ the function
(\ref{eq.13}) is more sensitive to different noise levels.
\begin{table}
\caption{\label{tab:tab.2} Results of noise level estimation for
the Lorenz system with the dynamical noise.}
\begin{ruledtabular}

\begin{tabular}{cccc}

  System & $\%NTS$& $\sigma$& estimated $\sigma$\\
\hline
  Lorenz &$11\%$& 1 & $1.19\pm 0.12$\\
  Lorenz &$11\%$& 1 & $1.17\pm 0.15$\\
  Lorenz &$11\%$& 1 & $1.14\pm 0.1$\\
  Lorenz &$11\%$& 1 & $1.15\pm 0.2$ \\
  Lorenz &$11\%$& 1 & $1.11\pm 0.18$ \\
  Lorenz &$11\%$& 1 & $1.09\pm 0.14$ \\

\end{tabular}
\end{ruledtabular}
\end{table}
\par
\begin{table}
\caption{\label{tab:tab.3} Results of noise level estimation for
systems with mixture of measurement and dynamical noise.}
\begin{ruledtabular}
\begin{tabular}{cccc}

  System &$\%NTS$& $\sigma$& estimated $\sigma$
  \\\hline
  Lorenz & $43\%$ & 4.06 & $4.56\pm 0.12$ \\
  Lorenz &$56\%$ & 5.93 & $5.34\pm 0.11$ \\
  Lorenz &35$\%$& 2.93 & $2.42\pm 0.12$\\
  Roessler &14$\%$& 2.82 & $1.97\pm 0.12$\\
  Roessler &$94\%$& 33.5 & $32\pm 0.75$ \\
  Roessler &$81\%$& 16.12 & $16\pm 0.71$ \\
 \end{tabular}
\end{ruledtabular}
\end{table}
\par
To verify our method in  a real experiment we have performed
analysis of data generated by a nonlinear electronic circuit. The
Chua circuit in the chaotic regime \cite{chua1,chua2} has been
used and we have added  a measurement noise to the outcoming
signal. The noise (white and Gaussian) has come from an electronic
noise generator. The results are presented in the table
\ref{tab:tab.4}. The first two rows correspond to $N=10000$ and
the rest to $N=1000$. In the case of a small noise level we can
not perform any estimation for a  small number of data, because
the noise is smaller than the average distance between nearest
neighbors. The estimation for $N=1000$ has taken a few minutes
\cite{computer}.
\begin{table}
\caption{\label{tab:tab.4} Results of noise level estimation for
the Chua circuit with the measurement noise.}
\begin{ruledtabular}

\begin{tabular}{ccc}

  $\%NTS$& $\sigma$ [mV]& estimated
  $\sigma$[mV]
  \\\hline
  $0\%$& 0  & $0.15\pm 0.015$ \\
  $3.1\%$& 30.4  & $29.6\pm 0.3$ \\
  $6.2\%$& 60.8 & $61.3\pm 8$\\
  $12.3\%$& 121.7 & $116\pm 8$\\
  $24.9\%$& 243.4 & $223\pm 13$ \\
  $28.3\%$& 304 & $380\pm 9$ \\
  $46.1\%$& 486 & $499\pm 20$\\
  $73.7\%$& 973 & $1109\pm 52$\\
  $90.6\%$& 1520 & $1537\pm 17$\\
  $96.5\%$& 2120 & $2042\pm 38$\\
\end{tabular}
\end{ruledtabular}
\end{table}

\section{Conclusions}
\par In conclusion we have developed a new method of the noise level estimation from time series.
The method makes use of the functional dependence of the coarse
grained entropy $K_2\lb\eps\rb$ on the threshold $\eps$. It
appears that the peculiar shape of this entropy $K_2\lb\eps\rb$
depends on the standard deviation of the noise $\sigma$ so a
simple function fitting can be applied to find the noise level.
The process of noise estimation can be done easily without
assuming input parameters and can be programmed in such a way that
the algorithm makes all steps automatically. When the length of
the time series $N<5000$ the whole evaluation procedure takes a
few minutes \cite{computer}. The method has no limitations
regarding a noise level and a kind of noise so one can evaluate
very high noise levels and a dynamical noise as well. We have
verified the validity of our method by applying it to estimate the
noise level in several  chaotic systems and in the Chua electronic
circuit.
\begin{acknowledgments} We are thankful to Prof. Hartmut Benner and to Prof. Dirk
Helbing for their hospitality during our stays at Darmstadt
University of Technology and at the Technical University of
Dresden. This work was in part supported by the special Program
\textit{Dynamics of Complex Systems} of Warsaw University of
Technology and by the Quandt Foundation of the ALTANA AG.
\end{acknowledgments}


\begin{thebibliography}{01}
\bibitem{kantzschreiber} H. Kantz and T. Schreiber, \textit{Nonlinear Time Series Analysis} (Cambridge University Press, Cambridge, 1997).
\bibitem{abarbanel} H.D.I. Abarbanel, \textit{Analysis of Observed Chaotic Data} (Springer, New York, 1996).
\bibitem{Schreiber1} T. Schreiber, Phys. Rev. E \textbf{48(1)},13(4) (1993).
\bibitem{Farmer} J. D. Farmer and J.J. Sidorowich, Physica D \textbf{47}, 373-392 (1991).
\bibitem{Diks} C. Diks, Phys. Rev. E \textbf{53(5)},4263(4) (1996).
\bibitem{Yu} Dejin Yu, M. Small, R.G. Harrison and C. Diks, Phys. Rev. E \textbf{61(4)},3750(7) (2000).
\bibitem{Hsu} R. Cawley and Guan-Hsong Hsu, Phys. Rev. A \textbf{46(6)}, 3057 (1992).
\bibitem{Oltmans}H. Oltmans and P. J.T.Verheijen, Phys. Rev. E \textbf{56(1)},1160(11) (1997).
\bibitem{Eckmann} J-P. Eckmann, S Kamphorst and D. Ruelle, Europhys. Lett. \textbf{4}, 973-977 (1987).
\bibitem{Zbilut} L.L. Trulla, A. Giuliani, J.P. Zbilut and C.L. Webber Jr., Phys. Lett. A \textbf{223}, 255-260 (1996).
\bibitem{manetti} C. Manetti, M. A. Ceruso, A. Giuliani, C. L. Webber Jr. and J. P. Zbilut, Phys. Rev. E \textbf{59}, 992-998 (1999).
\bibitem{giuliani} J. P. Zbilut, A. Giuliani and C. L. Webber Jr., Phys. Lett. A \textbf{246}, 122 (1998).
\bibitem{Holyst} J.A. Ho{\l}yst, M. \.Zebrowska and K. Urbanowicz, European Physical Journal B \textbf{20}, 531-535 (2001).
\bibitem{Atay} F.M. Atay and Y. Altintas, Phys. Rev. E \textbf{59(6)}, 6593(6) (1999).
\bibitem{Chai} J.M. Chai, B.H. Bae and S.Y. Kim, Phys. Lett. A \textbf{263}, 299-306 (1999).
\bibitem{Matassini} L. Matassini, H. Kantz, J. Ho{\l}yst and R. Hegger, Phys. Rev. E \textbf{65},021102(2002).
\bibitem{Hegger} R. Hegger, H. Kantz, L. Matassini and T. Schreiber, Phys. Rev. Lett. \textbf{84},
4092 (2000).
\bibitem{Faure} P. Faure and H. Korn, Physica D \textbf{122}, 265-279 (1998).
\bibitem{Shan} The coarse grained Kolmogorov-Sinai entropy is
related to so-called $\epsilon$-entropy introduced by Shannon, see
C.E. Shannon, Bell Syst. Techn. J. \textbf{27}, 379 and 623
(1948).
\bibitem{wang} P. Gaspard and Xiao-Jing Wang, Phys. Rep.
\textbf{235}, 291-343 (1993).
\bibitem{zyczkowski} A. Ostruszka, P. Pako{\.n}ski, W. S{\l}omczy{\.n}ski and K.
{\.Z}yczkowski, Phys. Rev. E \textbf{62}, 2018-2029 (2000).
\bibitem{cohen} A. Cohen and I. Procaccia, Phys. Rev. A
\textbf{31}, 1872 (1985).
\bibitem{computer} A processor {\textit Celeron} $400$MHz has been used for numerical
calculations.
\bibitem{Schreiber} P. Grassberger, R. Hegger, H. Kantz, C.
Schaffrath and T. Schreiber, Chaos \textbf{3(2)},127 (1993).
\bibitem{Saure} E.J. Kostelich and T. Schreiber, Phys. Rev. E \textbf{48(3)},1752 (1993).
\bibitem{Proccacia} P. Grassberger and I. Procaccia, Phys. Rev. A \textbf{28}, 2591 (1983).
\bibitem{Benettin} G. Benettin, L. Galgani and J.M. Strelcyn, Phys. Rev. A \textbf{14(6)}, 2338 (1976).
\bibitem{Schuster} K.Pawelzik and H.G. Schuster, Phys. Rev. A \textbf{35}, 481 (1987).
\bibitem{Grassberger} P. Grassberger and I. Procaccia, Phys. Rev. Lett. \textbf{50(5)}, 346 (1983).
\bibitem{Ghez} J.-M. Ghez and S. Vaienti, Nonlinearity \textbf{5},777-790 (1992).
\bibitem{boffetta} G. Boffetta, M. Cencini, M. Falcioni and A. Vulpiani, Physics Reports \textbf{356}, 367-474 (2002).
\bibitem{cencini} M. Cencini, M. Falcioni, E. Olbrich, H. Kantz and A. Vulpiani, Phys. Rev. E \textbf{62(1)},427(11) (2000).
\bibitem{NR} W. H. Press, S. A. Teukolsky, W. T. Vetterling and B. P. Flannery, \textit{Numerical recipies in C} (Cambridge
Univesrity Press, second edition, 1992).
\bibitem{Jaeger} L. Jaeger and H. Kantz, Physica D \textbf{105}, 79-96 (1997).
\bibitem{chua1} Shuxian Wu, Proceedings of the IEEE, vol. \textbf{75}, No. 8 (1987).
\bibitem{chua2} L.O Chua and G.-N. Lin, IEEE Transactions on Circuits and Systems \textbf{37(7)},885-902 (1990).

\end{thebibliography}

\end{document}